\documentclass{article}
\usepackage{spconf,amsmath,graphicx,balance}

\usepackage{todonotes}
\usepackage{tikz}
\usetikzlibrary{spy}
\usepackage{subcaption}
\usepackage{siunitx}

\usepackage{enumitem}
\setlist{nosep} 

\usepackage{subcaption}

\newcommand{\etal}{\textit{et al.}}

\usepackage{changes}
\definechangesauthor[name={Merlin Nau}, color=violet]{mn}

\usepackage{changes}
\definechangesauthor[name={Florian Schiffers}, color=orange]{fs}

\usepackage{changes}
\definechangesauthor[name={Oliver Cocssairt}, color=blue]{os}

\usepackage{changes}
\definechangesauthor[name={Florian willomitzer}, color=red]{fw}

\title{SkinScan: Low-cost 3D-scanning for dermatologic diagnosis and documentation}
\makeatletter

\def\@name{\em{Merlin A. Nau$^{1,2*}$, Florian Schiffers$^1$, Yunhao Li$^1$, Bingjie Xu$^1$, Andreas Maier$^2$}
\\
\emph{Jack Tumblin$^1$, Marc Walton$^4$, Aggelos K. Katsaggelos$^{1,3,5}$, Florian Willomitzer$^3$, Oliver Cossairt$^{1,3}$}
\thanks{Thanks to Northwestern Alumnae for funding.}
\thanks{\copyright 2021 IEEE. Personal use of this material is permitted. Permission from IEEE must be obtained for all other uses, in any current or future media, including reprinting/republishing this material for advertising or promotional purposes, creating new collective works, for resale or redistribution to servers or lists, or reuse of any copyrighted component of this work in other works.}
}

\address{
$^1$Department of Computer Science, Northwestern University, Evanston, USA \\
$^2$Pattern Recognition Lab, Friedrich-Alexander-Universit{\"a}t Erlangen-N{\"u}rnberg, Germany\\
$^3$Department of Electrical and Computer Engineering, Northwestern University, Evanston, USA\\
$^4$Center for Scientific Studies in the Arts, Northwestern University, Evanston, USA\\
$^5$Department of Radiology, Northwestern University, Chicago, USA\\
{\small $^*$ merlin.nau@fau.de}
} 

\begin{document}
%
%
\maketitle
\begin{abstract}

The utilization of computational photography becomes increasingly essential in the medical field. 
Today, imaging techniques for dermatology range from two-dimensional (2D) color imagery with a mobile device to professional clinical imaging systems measuring additional detailed three-dimensional (3D) data.
The latter are commonly expensive and not accessible to a broad audience. 
In this work, we propose a novel system and software framework that relies only on low-cost (and even mobile) commodity devices present in every household to measure detailed 3D information of the human skin with a 3D-gradient-illumination-based method. 
We believe that our system has great potential for early-stage diagnosis and monitoring of skin diseases, especially in vastly populated or underdeveloped areas.
\end{abstract}
\section{Introduction}
\label{sec:intro}


%
%

%
%


3D scanning provides access to a plethora of useful features in the diagnosis and documentation of skin diseases~\cite{rahman2016early}.
For example, skin cancer is the most common cancer in the United States developed by every fifth person throughout their lifetime~\cite{stern2010prevalence}.
While skin cancer is treatable, early detection is essential for a successful prognosis.
Therefore, routine screenings are vital to ensure an early-stage diagnosis and therapy~\cite{mohan2014advanced}.

Generally, the most important evaluation for a correct diagnosis of skin cancer is based on the \textit{ABCDE}~\cite{rigel2005abcde} guided rule assessing \textbf{a}symmetries of the lesion shape, \textbf{b}order irregularities, \textbf{c}olor variegation, the lesion’s \textbf{d}iameter and the \textbf{e}volution of shape and color.
Currently, it is most common to evaluate the above criteria solely based on 2D image information. 
However, since lesions can also grow in height, this evaluation can be significantly improved with additional access to 3D topographic information of the skin ~\cite{ares2014handheld} (in fact, melanomas at a young age are misdiagnosed by up to 40\%, in comparison with pigmented lesions~\cite{ferrari2005does}).
More to the point, surface orientation on skin cancer is a feature that can support the distinction between malign and benign tissue ~\cite{she2013lesion}.
Although increasingly used by professional physicians in their diagnosis and documentation, clinical 3D imaging techniques to measure human skin like optical coherence tomography~\cite{mogensen2009oct} or laser-based scanners~\cite{zenteno2017volume}, are tied to expensive hardware, are not portable, and hence are not accessible to a broad audience.

\begin{figure}[tb]
    \centering
    \includegraphics[width=\linewidth]{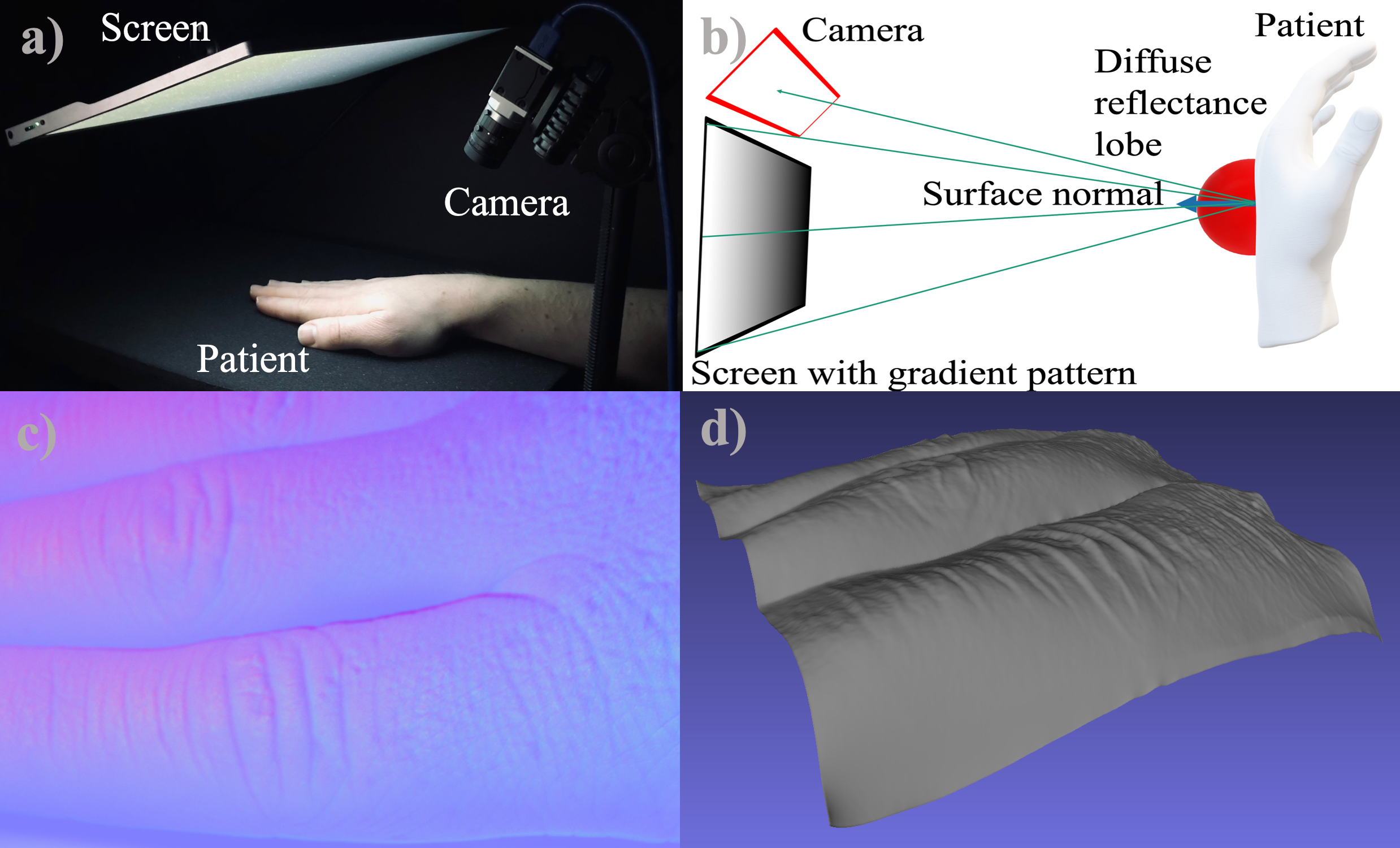}/
    \caption{a) Experimental setup with a vision camera and a lightweight screen. b) Illustration of gradient illumination. c) The recovered normal map. d) 3D reconstruction}
    \label{fig:setup}
\end{figure}

In this work, we ask the question:
Can routine examinations of skin disease be augmented with low-cost but precise 3D imaging techniques to improve assessment, monitoring, and diagnosis effectiveness?
Mobile applications for phones or tablets that allow the exchange of images between the medical doctor and patient are broadly available ~\cite{trettel2018telemedicine}.
However, these apps only allow for patient-driven image capture of  2D RGB images.
Currently patients cannot capture and transfer 3D information of their skin abnormalities to a medical doctor for further evaluation. 
This feature becomes even more critical during the time of the COVID pandemic.
The main advantages of 3D over 2D imaging are that the data is invariant to the translation and rotation of the object, surface texture, and external illumination.
These advantages can significantly improve the repeatability of patient-driven measurements, enabling a better dermatological evaluation across different skin tones, pigmentation variations, as well as, skin conditions or tattoos.

While some success has been made in developing state-of-the-art artificial intelligence classification systems for skin diseases from 2D images alone, the initial attempts also leave room for significant improvement~\cite{kinyanjui2020fairness}.
In particular, additional depth information could prove of benefit for Deep-Learning (DL) techniques by enriching the feature space.

Previously, several approaches using structure-from-motion or triangulation (``structured ligh'') have been explored for mobile wound documentation and measurement~\cite{kumar2019comparison}.
However, most open source proposed methods turned out to be too imprecise to provide detailed structural images of skin topography~\cite{kumar2019comparison}. 
Active triangulation techniques perform significantly better but require a suitable and expensive sensor~\cite{ares2014handheld}.

In this paper, we propose to use mobile gradient illumination~\cite{riviere2016mobile} to build a low-cost measurement system for dermatological evaluation of skin.
In particular, our main contributions are
\begin{enumerate}
    \item A method for inexpensive 3D capture of skin using commodity hardware (like screens and webcams, or tablets)
    \item Preliminary results of 3D skin capture using the proposed method
    \item Development of an open-source framework for gradient illumination-based 3D capture of skin.
\end{enumerate}
While this paper's focus is on the analysis and documentation in medical 3D imaging, the method we have developed could also have much broader applications.
 For instance, industrial object inspection or conservation in cultural heritage poses exciting scenarios for a mobile and low-cost 3D method \cite{willomitzer2020hand}.
\vspace{-5 mm}
\section{3D Imaging of Human Skin}

%
%
%

3D imaging poses a set of challenges with no universal solution. 
Achieving reliable measurements for different materials at arbitrary spatial and temporal resolutions is not feasible.
%
%
Passive methods, such as \textit{Passive stereo} ~\cite{lazaros2008review}, \textit{Structure from Motion} \cite{schonberger2016structure}, or even single-frame DL-based depth estimation \cite{liu2015deep} are not able to resolve fine details in 3D and are not feasible for our proposed task.

%

Active techniques, such as time-of-flight~\cite{kolb2010time} or active triangulation (``structured-light'')~\cite{willomitzer2017single}, employ an additional light-source or projector to encode depth-information into the measurement. While active triangulation techniques would provide the required measurement precision on human skin~\cite{willomitzer2017single}, the hardware necessary to perform the respective measurement is still rarely found in consumer devices.
%
%
Another active method able to measure small surface details is \textit{Photometric Stereo} (PS)~\cite{sohaib2013vivo}.
%
%
%
%
%

%
Here, multiple light sources are placed around the diffuse object, capturing images with each light source switched on and the remaining lights off.
Compared to 3D estimations from shading in a single frame, the illumination from various directions supports the reliability of results. 
%
%
To be concise, the amount of reflected light is dependent on the light source positions relative to the object. 
Thus, one can resolve the surface orientation.
Debevec~\etal~\cite{debevec2012light} extended PS by the idea of placing multiple light sources and cameras in a \textit{Light stage} surrounding the object of interest.
However, the complicated and bulky setup makes it difficult to apply it in a mobile setting.
%
%
%
%
%
%
%
%
%
%

%
%
Ma \etal~\cite{ma2007rapid} showed that a screen displaying intensity gradients could replace multiple light sources. The resulting  \textit{Gradient Illumination} (GI) approach can recover accurate 3D information by using a single camera with a polarization lens and a half-spherical-shaped light source.
In 2016, Riviere \etal~\cite{riviere2016mobile} proposed a mobile extension by using gradients displayed on the tablet screen to recover the surface normals for specular and diffuse reflection by placing a polarization lens on the front camera.
All these so-called ``reflectance-based'' methods are highly dependent on the reflective properties of the objects of interest.
For specular objects, Willomitzer \etal~\cite{willomitzer2020hand} proposed the idea of utilizing a tablet to perform deflectometry measurements using the screen and front camera. However, this system is not tailored to measure rather diffuse objects like human skin. This paper's method offers substantially improved results on human skin.

\begin{figure}[tb]
    \centering
    \includegraphics[width=\linewidth]{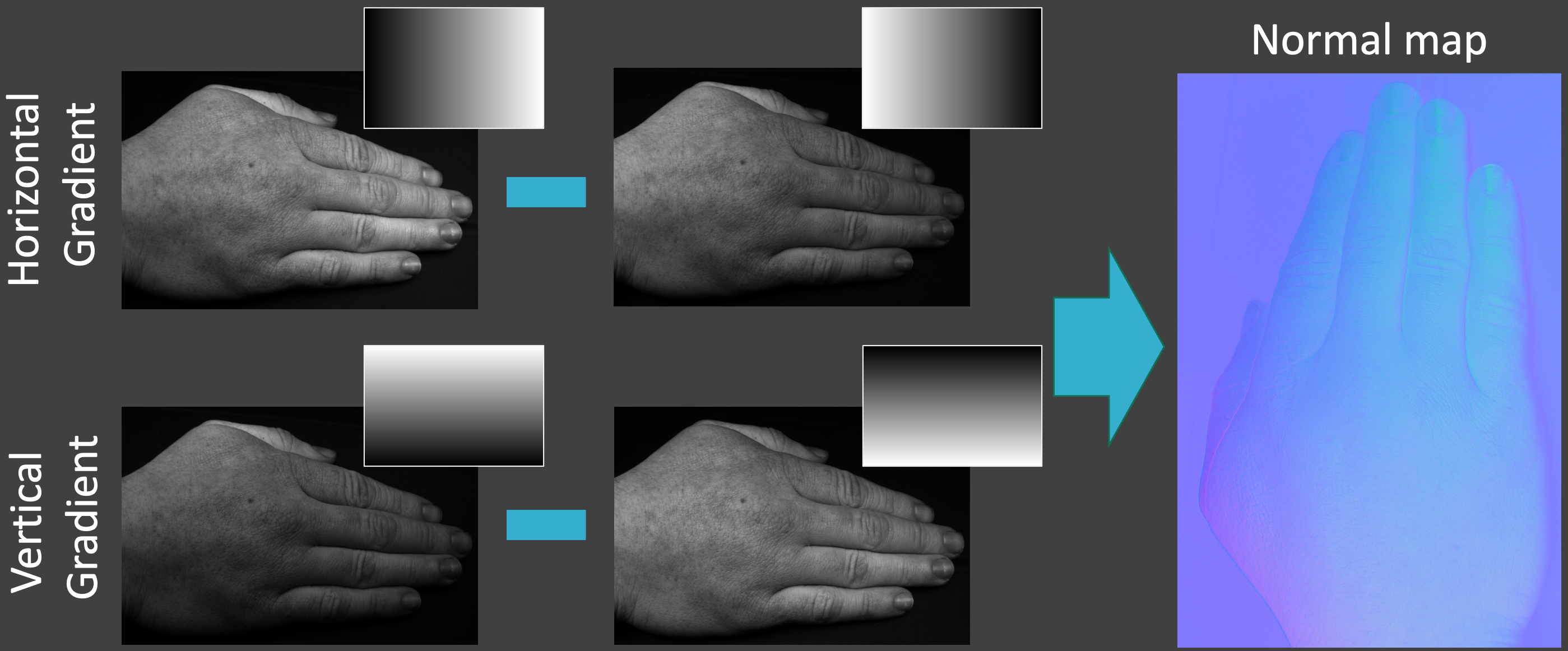}
    \caption{Illustration of the gradient illumination procedure. Images are acquired under each gradient pattern in horizontal and vertical direction. From there the directional images are subtracted, forming $x$ and $y$ component of the normal map.}
    \label{fig:graphical_abstract}
\end{figure}
\vspace{-3 mm}

\section{Methods}
\vspace{-2 mm}
\label{sec:format}

%
%

%
For our desired application of measuring human skin, we identify gradient illumination as a suitable measurement technique.
A depth map can be obtained from the measured surface normal data by integration algorithms, such as, the Frankot-Chellapa algorithm~\cite{frankot1988method}.
Gradient illumination employs an electronic display where each pixel can be considered as an individual light-source.

Our experimental setup consists of only a programmable screen and a camera (see Fig.~\ref{fig:setup}) and, in principle, can be implemented with just a single mobile device (\textit{e.g.}, smartphone or tablet) according to the solution presented in \cite{willomitzer2020hand}. 
When imaging objects with Lambertian reflectance, one can define the observed radiance $r$ from a viewing position $\vec{v}$ (\textit{e.g.}, a camera ray in Fig. 2h) as an integration over a product of the incident illumination direction vector~$\vec{\omega}$ created by the gradient patterns  $P(\vec{\omega})_{x,y}$ programmed on the screen, and the bidirectional reflectance distribution function (BRDF).
For diffuse objects the BRDF $R$ is approximated as the product of the diffuse albedo $\rho_d$ and the foreshortening factor, dependent on $\vec{\omega}$ and the surface normal direction $\vec{n} = (n_x, n_y, n_z)^T $:
\begin{equation}
R(\vec{\omega}, \vec{n}) = \rho_d \cdot  \text{max}(\vec{\omega} \cdot \vec{n}, 0)
\end{equation}
\begin{equation}
\label{eqn:eq1}
r(\vec{v})= \int_{\Omega} P_i(\vec{\omega}) R(\vec{\omega}, \vec{n}) d \vec{\omega}
\end{equation}

\begin{figure}[tb]

\begin{subfigure}[b]{0.4\linewidth}
    \centering
    \includegraphics[width=\linewidth,angle=90]{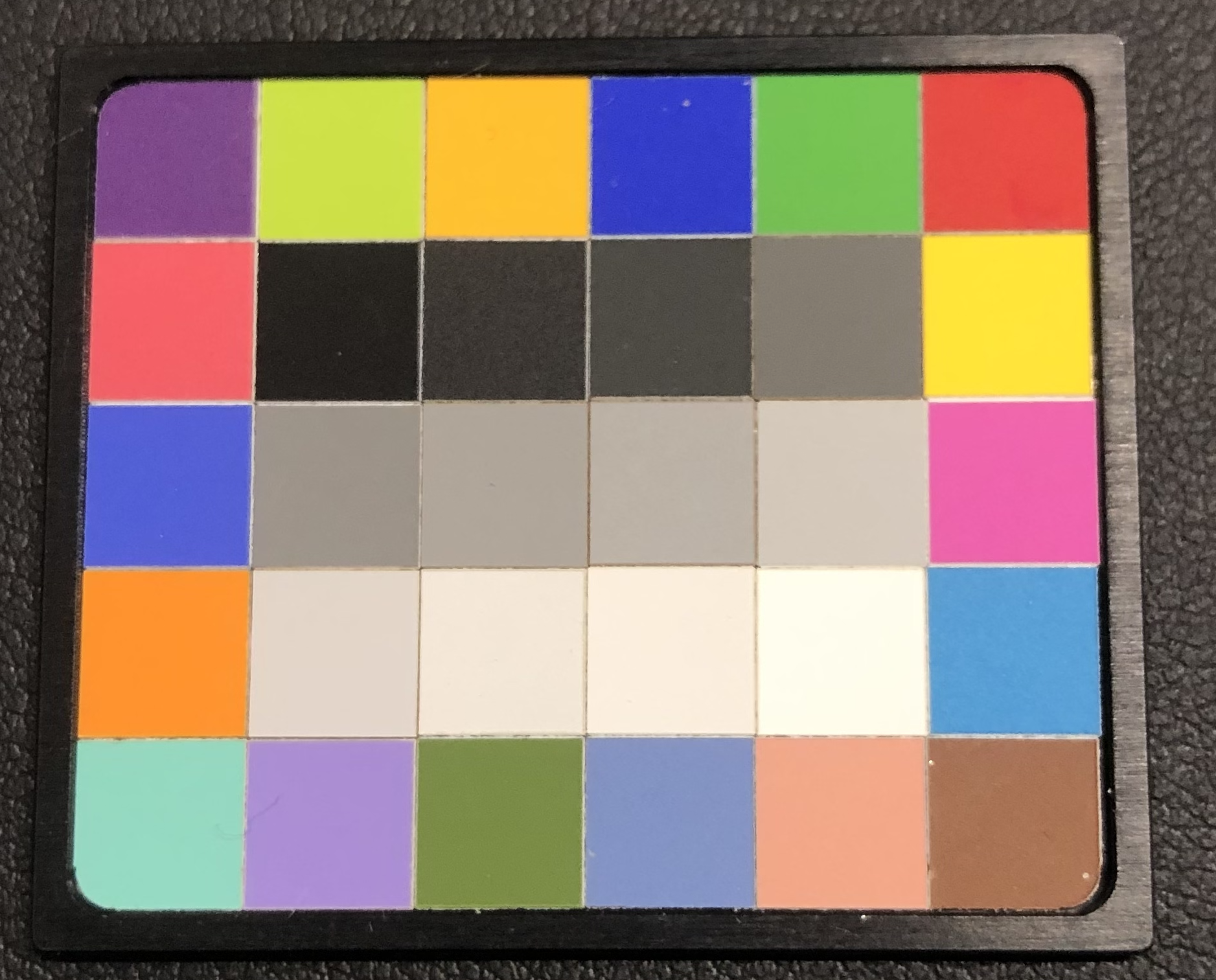}
    \caption{Colorchecker}
    \label{fig:colorchecker}
\end{subfigure}
\hfill
\begin{subfigure}[b]{0.59\linewidth}
    \centering
    \includegraphics[width=\linewidth]{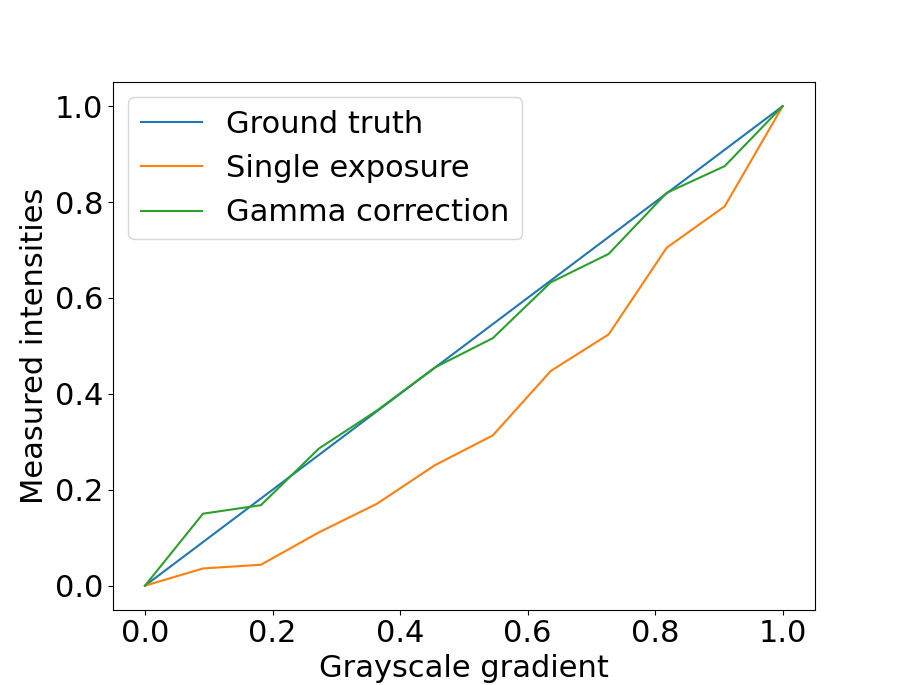}
    \caption{Calibration Curve}
    \label{fig:calibration}
\end{subfigure}

\caption{Radiometric calibration using a color chart.}

\end{figure}

Because a screen can only project gradients along two directions, \textit{e.g.}, $x$ and $y$, we assume $\vec{n}$ to be aligned with $\vec{z} = [0,0,1]$, which coincides with the centroid of a diffuse reflectance lobe of a flat surface pointing towards the screen.
By integrating Eq.~\ref{eqn:eq1} the radiance is resolved according to \cite{ma2007rapid} as
\begin{equation}
\label{eqn:eq2}
r_x(\vec{v})=n_x(\frac{2 \pi \rho_d}{3})
\end{equation}

Compared to Ma \etal~\cite{ma2007rapid}, we do not integrate over the entire hemisphere. 
Instead, the integration is only performed in the $x$ and $y$ directions. 
Displaying the gradient patterns in both $x$ and $y$ directions for both $[0, 1]$ and $[1, 0]$, we  calculate $r_x, r_y$ and $\hat{r}_x$ $\hat{r}_y$, respectively.
We can resolve the $x$ and $y$ components of the normal by:
\begin{equation}
\label{eqn:eq3}
n_{x,y} = r_{x,y} - \hat{r}_{x,y}
\end{equation}
Finally, we estimate the z-component assuming a normal by:
\begin{equation}
\label{eqn:eq4}
n_z = \sqrt{1-n_x^2-n_y^2}
\end{equation}

\vspace{-2 mm}
\subsection*{Open source software framework}
\label{ssec:subhead1}

The methods presented in this paper are part of an open-source software framework that can be downloaded from \footnote[1]{https://github.com/merlzbert/SkinScan}. The framework is implemented in Python and has been tested on Ubuntu, macOS, and Windows.
To encourage individual adjustments and extensions, we subdivide our project in an object-oriented manner with base classes for components necessary in any structured light setup.
These include a camera, projector, image reconstruction, and calibration class.
Consequently, the classes for the specific cameras, projectors, and reconstruction and radiometric, intrinsic, and extrinsic calibration are inherited from these base classes, allowing specific adjustments.
Namely, we have adapted camera classes for \textit{Basler cameras}, the \textit{Pi-Cam} and \textit{Webcams}. 
The projection unit provides functionality for any display or projector.

Radiometric calibration is accomplished by placing a color chart next to the sample, as displayed in Fig.~\ref{fig:colorchecker}.
Consequently, the camera's intensity values are linearized by applying a gamma correction, see Fig.~\ref{fig:calibration}.
The last pattern displayed is always a constant illumination at maximum intensity, which allows for normalization across the frames.
Intrinsic calibration is implemented using the OpenCV tool kit~\cite{wang2010camera}.
Extrinsic calibration is accomplished by placing a mirror with ChArUco markers~\cite{an2018charuco} on the edges under the screen.
The pose of the mirror is retrieved from the ChArUco markers.
Further, a checkerboard pattern is displayed on the screen.
Consequently, the camera captures the reflection of the pattern in the mirror.
Thus rotation and translation between camera, object, and screen are retrieved.
The components are then used in dedicated calibration and capture sessions, yielding the calibration data, normal maps, integrated depth maps, point clouds, and color images.
The results are bundled and saved in a folder. 
\vspace{-2 mm}

\section{Results}
\label{sec:pagestyle}
%
%

In this section, we present reconstructed surface normal maps from different objects.
Our experiments' setup consists of a 15 inch LCD screen and a \textit{Basler ace 2} grayscale surface camera, both steadily mounted on a monitor arm and tripod, facing the object of interest.
A radiometric calibration has to be employed to ensure consistency of the computed normal orientations over a surface with color variance.
Therefore a matte \textit{Color Gauge Micro} chart, containing linearly reflective grayscale tiles, is used to correct the captured intensity values.
The intensity correction function to linearise the observed grayscale values is shown in Fig.~\ref{fig:calibration}.

\begin{figure}[tb]

\begin{subfigure}[b]{\linewidth}
\begin{tikzpicture}[spy using outlines={magnification=3.5,size=1.2cm, connect spies}]
\node {%
        \includegraphics[width=0.46\linewidth]{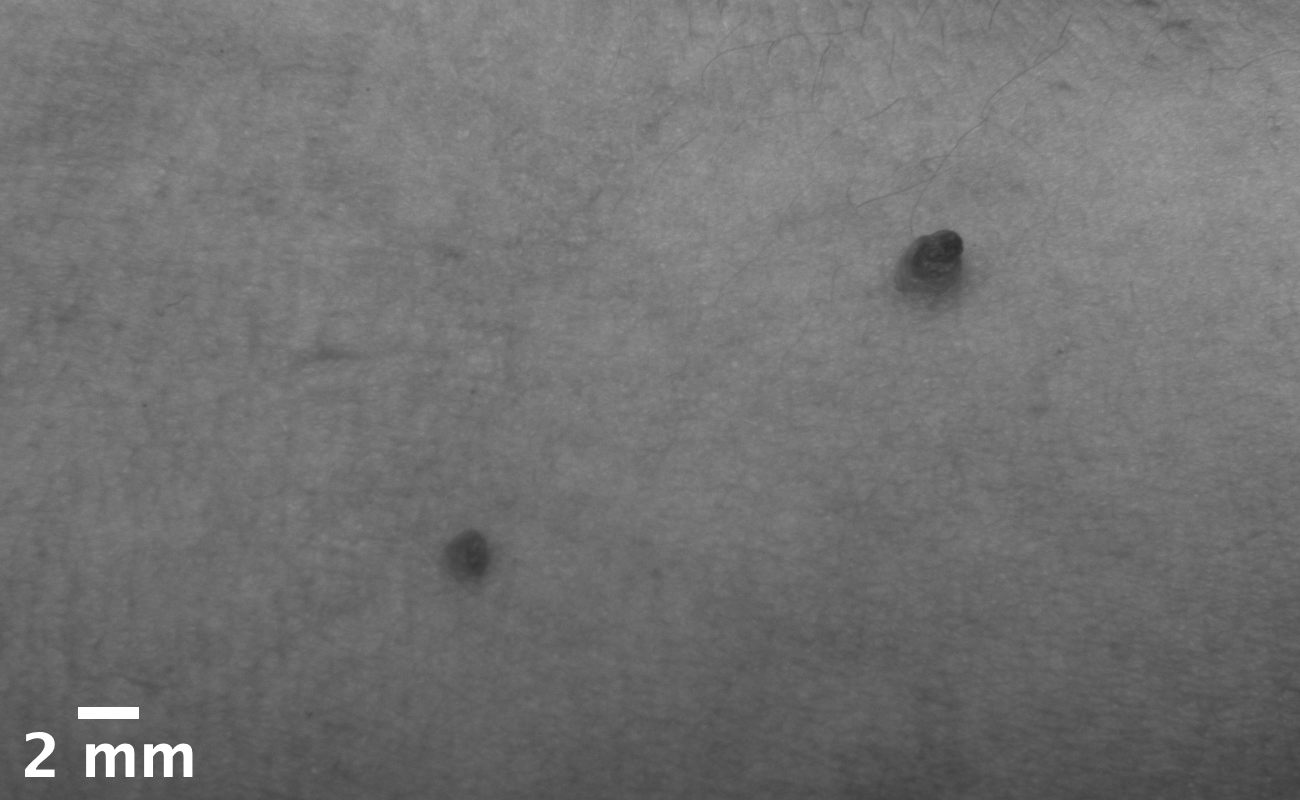}%
        };
    \node[anchor=west, text=white] at (-2.0, 1.0) {a)};
    \spy [blue, every spy on node/.append style={thick}] on (0.84, 0.4) in node [left] at (-0.3,0.55);
    \spy [red, every spy on node/.append style={thick}] on (-0.56, -0.5) in node [left] at (1.9,-0.55);

\end{tikzpicture}
\begin{tikzpicture}[spy using outlines={magnification=3.5,size=1.2cm, connect spies}]
\node {%
        \includegraphics[width=0.46\linewidth]{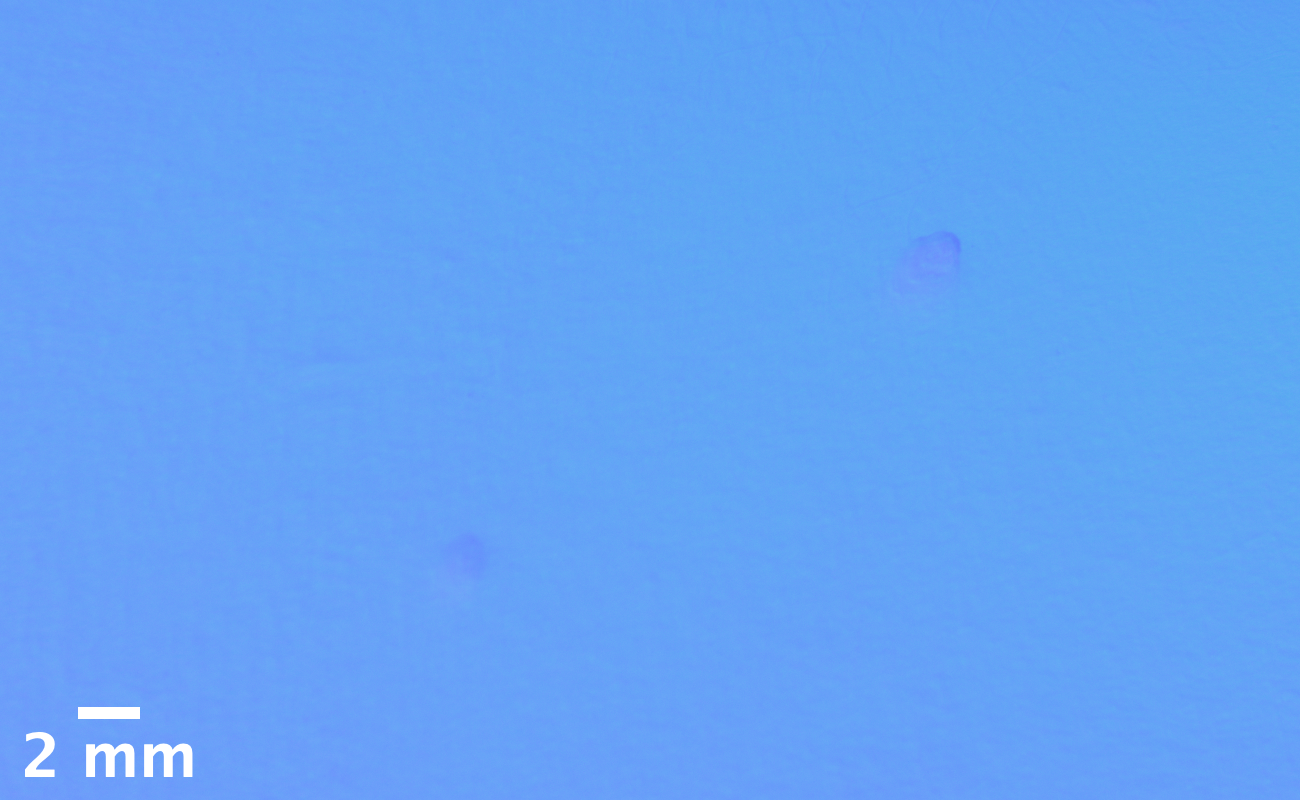}%
        };
\node[anchor=west, text=white] at (-2.0, 1.0) {b)};
\spy [blue, every spy on node/.append style={thick}] on (0.84, 0.4) in node [left] at (-0.3,0.55);
\spy [red, every spy on node/.append style={thick}] on (-0.56, -0.5) in node [left] at (1.9,-0.55);
\end{tikzpicture}
\end{subfigure}

\begin{subfigure}[b]{\linewidth}
\begin{tikzpicture}[spy using outlines={magnification=3.5,size=1cm, connect spies}]
\node {\includegraphics[width=0.46\linewidth]{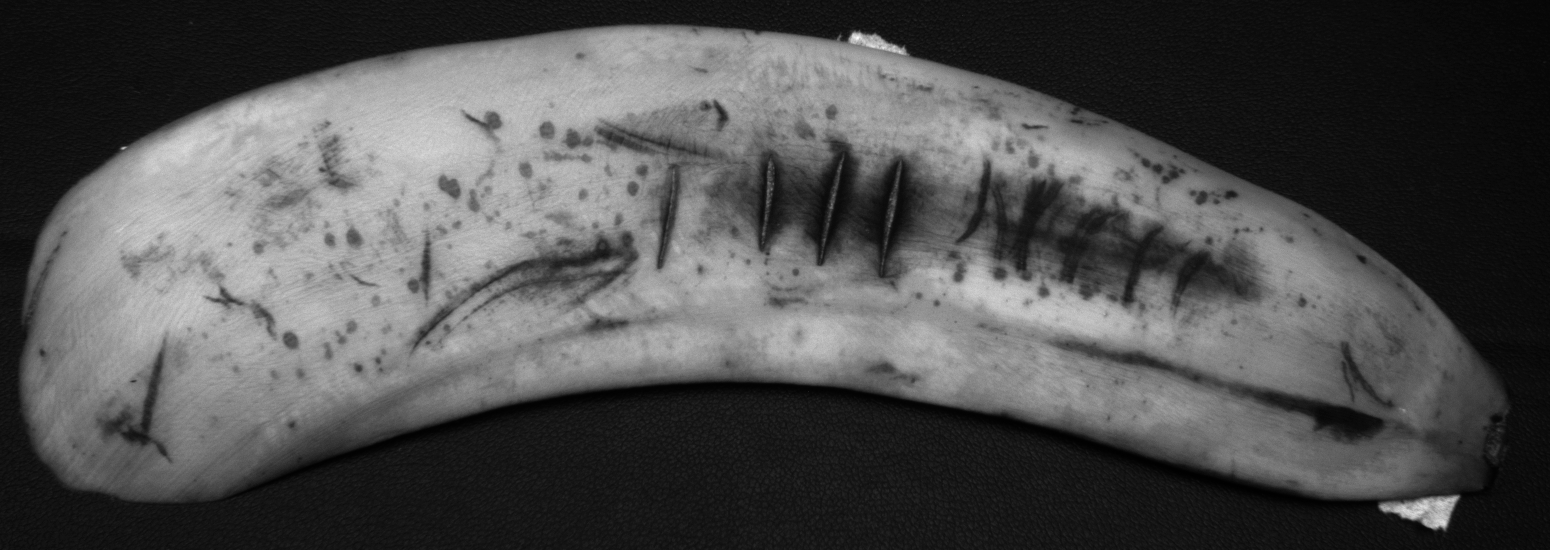}};
\node[anchor=west, text=white] at (-2.0, 0.5) {c)};
\spy [blue, every spy on node/.append style={thick}] on (0.05,0.2) in node [left] at (-0.5,0.0);
\spy [red, every spy on node/.append style={thick}] on (0.6, 0.2) in node [left] at (1.8,0.0);
\end{tikzpicture}    
\begin{tikzpicture}[spy using outlines={magnification=3.5,size=1cm, connect spies}]
\node {\includegraphics[width=0.46\linewidth]{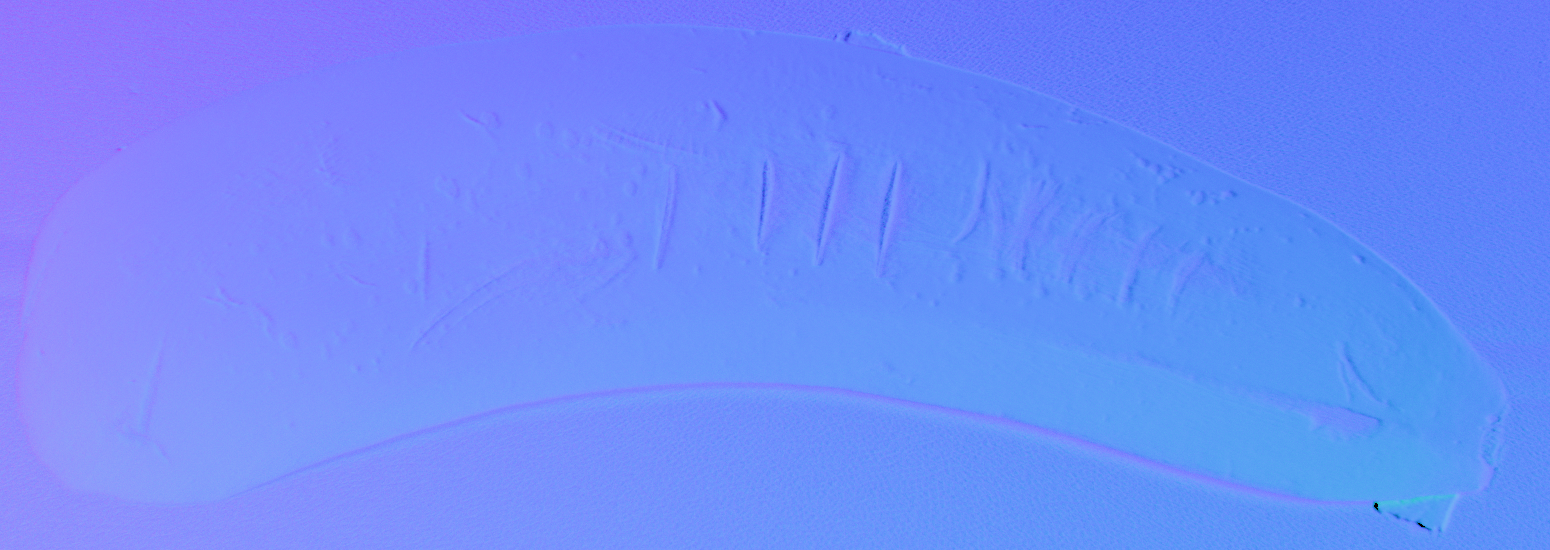}};
\node[anchor=west, text=white] at (-2.0, 0.5) {d)};
\spy [blue, every spy on node/.append style={thick}] on (0.05,0.2) in node [left] at (-0.5,0.0);
\spy [red, every spy on node/.append style={thick}] on (0.6, 0.2) in node [left] at (1.8,0.0);

\end{tikzpicture}
\end{subfigure}

\begin{subfigure}[t]{\linewidth}
\begin{tikzpicture}
    \node{%
        \includegraphics[height=3.7cm]{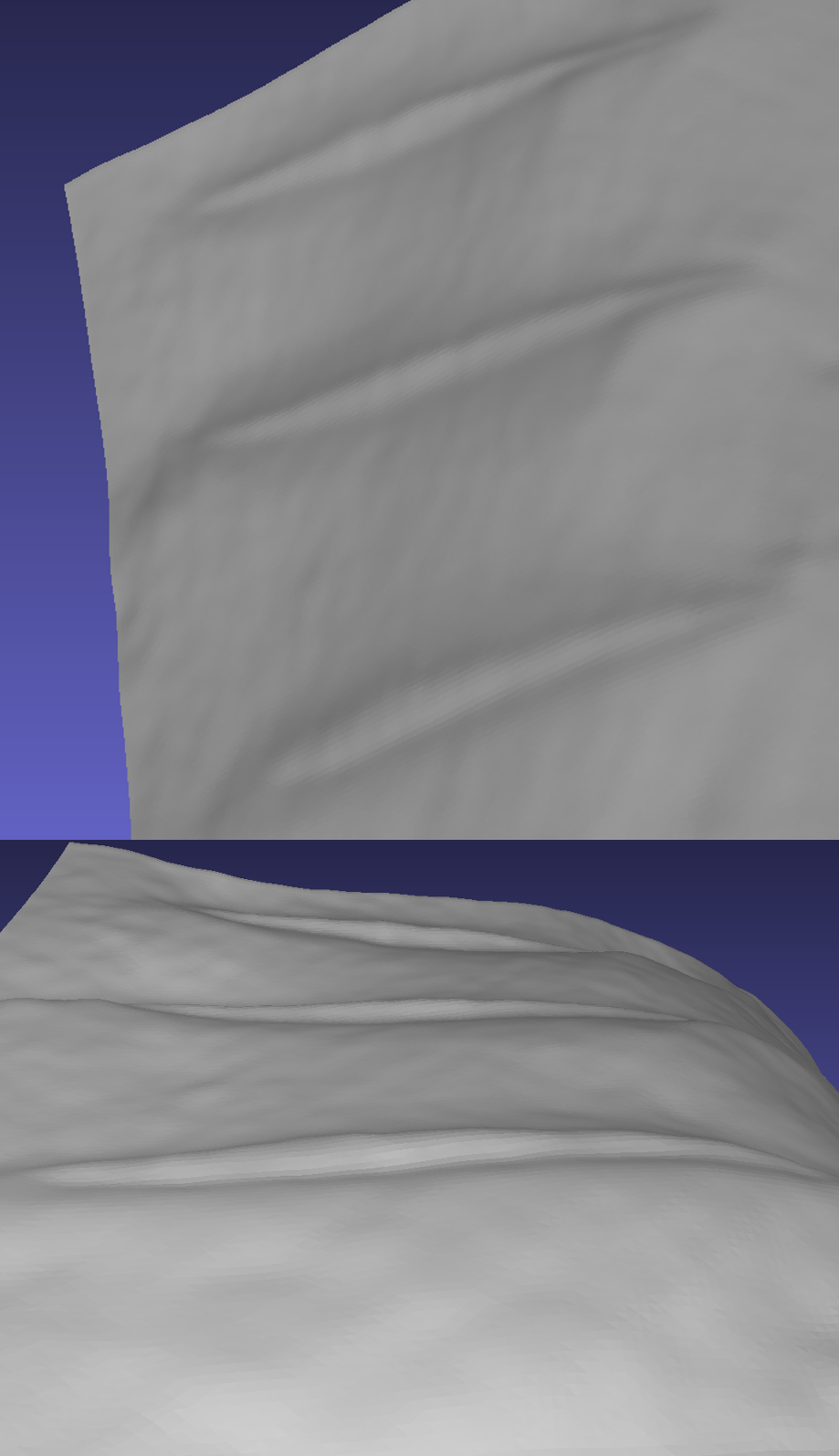}
        };
    \node[anchor=west, text=white] at (-1.1, 1.65) {e)};
\end{tikzpicture}
\begin{tikzpicture}
    \node{%
        \includegraphics[height=3.7cm]{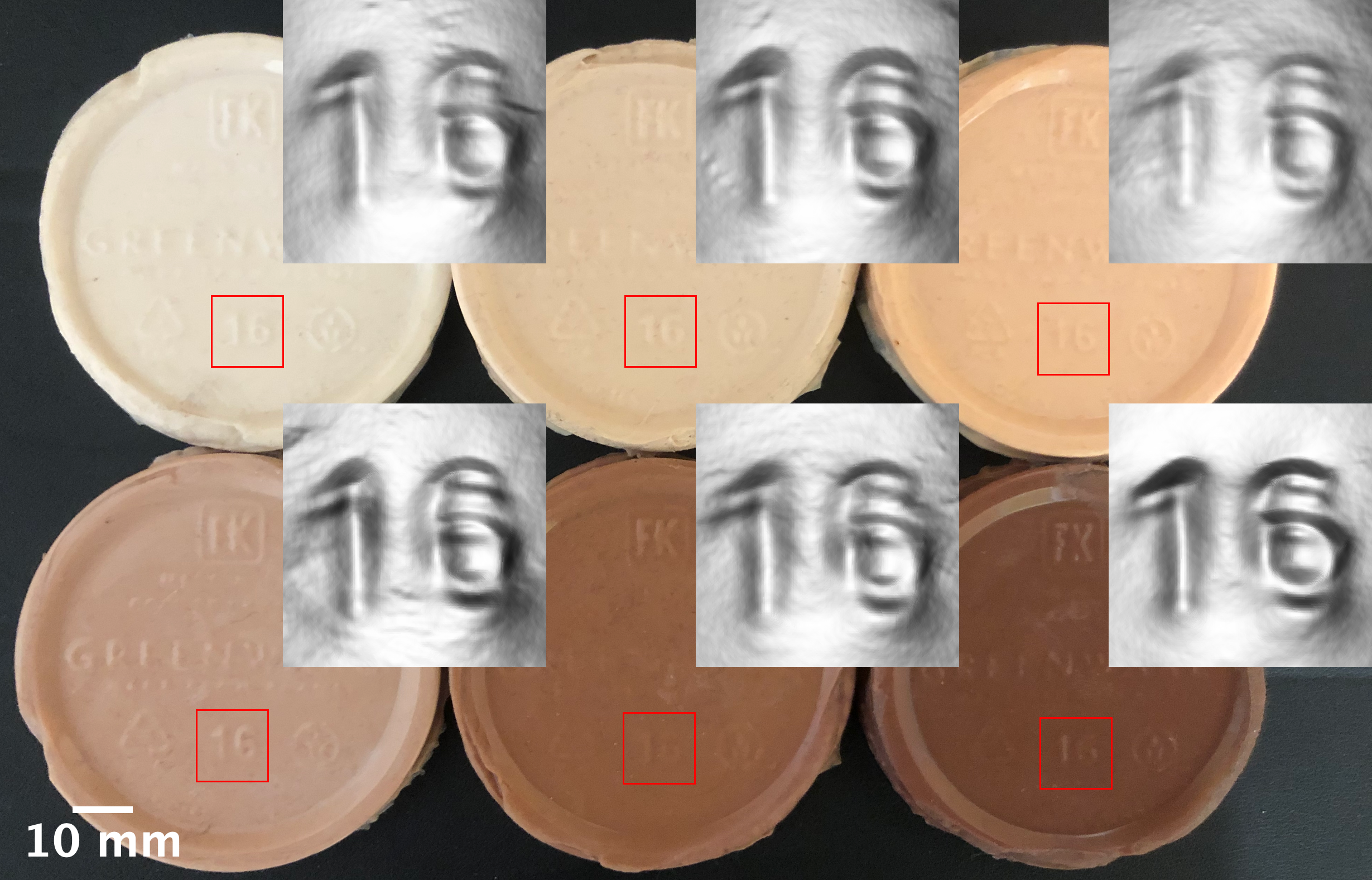}%
        };
    \node[anchor=west, text=white] at (-2.9, 1.65) {f)};
\end{tikzpicture}
\end{subfigure}


%



\caption{a) Banana peel with cuts and dents. b) Surface normal map the banana peel. c) Skin section containing two small moles. d) Surface normal map of skin section. e) 3D reconstruction of the cuts in the banana peel. f) Silicone skin replicas and their 3D reconstruction of the number `16'.}
\label{fig:reconstruction_results}

\end{figure}
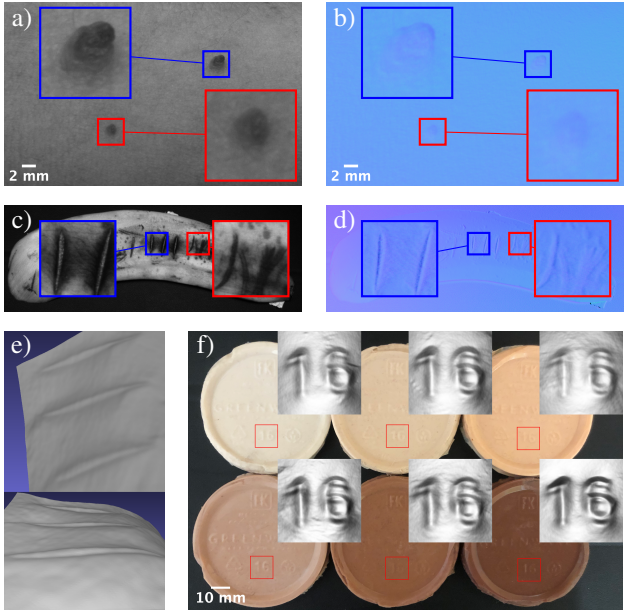

We present the acquired normal maps in a typical red, green and blue (RGB) fashion resembling normalized $x$, $y$, and $z$ components.
The normal maps alongside the grayscale albedo images are displayed in Fig.~\ref{fig:reconstruction_results}.
Moreover, the integration of the acquired surface normals yields a depth map.
Point clouds are created from the depth maps, which are saved in a mesh for visualization, see Fig.~\ref{fig:reconstruction_results} and Fig.~\ref{fig:setup}.
The texture on the point cloud is obtained by averaging the images captured under each gradient illumination.
As can be seen from the 3D rendering, in Fig.~\ref{fig:setup}, an accurate 3D representation of the folds and wrinkles of the skin has been captured, and the result is independent of environmental factors (\textit{e.g.}, ambient illumination, viewing direction, skin texture) so that the results are highly repeatable. 
As discussed, we postulate that such a method can significantly increase the reliability of assessment and monitoring of skin disease in remote health applications.

In addition to the results of a 3D-scan of a human hand shown in Fig.~\ref{fig:graphical_abstract}, we provide three more reconstructions showcasing that our proposed setup is a viable option for dermatologic diagnosis.
Firstly, we captured two moles arising from the surrounding tissue. 
The change of surface orientation, and therefore depth, is visible throughout the moles.
This advantage can be directly tied to the ABCDE rule.
Parameters of the rule may rely on color images only but can be significantly supported by shape estimation.
For example, color is not directly coupled to the depth of a mole. 
Furthermore, shape and color boundaries do not always coincide.
Finally, the color of an image is dependent on ambient lighting.
Secondly, we imaged a banana peel where the skin has darkened over several areas of the peel. 
We have further inserted small razor cuts as well as dents pushed in with a finger.
Specifically, it is essential to point out that the region around the cuts and dents has darkened.
Color variations can make the assessment of 3D structures from (2D) images hard.
As expected, the color variation is not visible in the normal map as the surface orientation remains unchanged. 
This shows that gradient illumination is independent of the surface texture albedo and indeed provides structural information.
In particular, the cuts and dents are visible.
While the reflectance properties of bananas are quite different from human skin, the example still shows that our method can factor the reflectance information from 3D shape features similar to what dermatological applications require.

%
To showcase another example of varying colors, we imitated different human skin tones with silicone used in the special effects industry. 
Consequently, silicone tones were molded in a plastic cup to recreate six human skin tones. 
The skin replicas and their corresponding 3D reconstructions of the embedded number `16' are shown in Fig.~\ref{fig:reconstruction_results}.
Although there is a slight variance between the samples, the number is resolved for all frames.
It is essential to mention that the arising number has a height of fewer than 0.5 millimeters. 
\vspace{-2 mm}

\section{DISCUSSION AND OUTLOOK}
\label{sec:typestyle}

We have proposed a method for mobile 3D imaging with low-cost components using gradient illumination for dermatological imaging.
The framework requires only a screen (\textit{e.g.}, a tablet or external monitor) and a camera (\textit{e.g.} the Pi-CAM) to provide high-quality 3D measurements of human skin or similar surfaces.
We have also developed an open-source framework to provide the software programming necessary for each component required by our method, including cameras, displays, calibration techniques, and image reconstruction algorithms.
%
%
%
%
Although we have focused on using our method for dermatological applications, its use may be much broader, \textit{e.g.}, in digital cultural heritage or biological imaging on-site.
%
%

While our method assumes diffuse reflectance, the reflection properties of skin across the entire human body and for various skin tones and ethnicities is much more complicated. As the next steps for our framework, we will incorporate approximate BRDF models and wavelength-dependent illumination to improve 3D reconstruction accuracy for arbitrary skin types.

Challenges that remain are the separation of diffuse and specular components on surfaces with varying reflective features without using a polarization lens. Specifically, this poses challenges for very shiny wounds with high fat or water content.

In future studies, we will evaluate the clinical use-case of our method with our clinical partners.
One aspect will be to develop machine-learning algorithms that improve the accuracy of the classical 2D-RGB skin disease classification algorithms by incorporating the structural 3D information provided by our method.
%
%
%

{ 
\ninept
\balance
\bibliographystyle{IEEEbib}
\bibliography{Template}
}

\end{document}